# Interaction between atoms and slow light: a waveguide-design study

Xiaorun Zang,[1] Jianji Yang,[1] Rémi Faggiani,[1] Christopher Gill,[2] Plamen G. Petrov,[2] Jean-Paul Hugonin,[3] Kevin Vynck,[1] Simon Bernon,[1] Philippe Bouyer,[1] Vincent Boyer,[2] and Philippe Lalanne[1,*]

[1]Laboratoire Photonique Numérique et Nanosciences (LP2N), UMR 5298, CNRS - IOGS - Univ. Bordeaux, 33400 Talence, France
[2]Univ. Birmingham, Sch Phys & Astron, Birmingham B15 2TT, W Midlands, England
[3]Laboratoire Charles Fabry, Institut d'Optique Graduate School, CNRS UMR 8501, Université Paris-Sud, 91127 Palaiseau Cedex, France.
*Corresponding author: philippe.lalanne@institutoptique.fr

**Abstract.** The emerging field of on-chip integration of nanophotonic devices and cold atoms offers extremely-strong and pure light-matter interaction schemes, which may have profound impact on quantum information science. In this context, a long-standing obstacle is to achieve strong interaction between single atoms and single photons, while at the same time trap atoms in vacuum at large separation distances from dielectric surfaces. In this work, we study new waveguide geometries that challenge these conflicting objectives. The designed photonic crystal waveguide is expected to offer a good compromise, which additionally allow for easy manipulation of atomic clouds around the structure, while being tolerant to fabrication imperfections.

## 1. Introduction

Strong atom-photon interaction, whereby a single photon can deterministically interact with a single atom, constitutes the basic building block of a number of quantum processing, simulation, and sensing schemes. Enhancing the atom-photon interaction requires increasing the mode electric field at the position of the atom. This can be done by reducing the transverse size of the light mode, for instance by guiding it in ultra-thin unclad optical fibers [1–3] or focusing it with high-numerical-aperture lenses [4,5]. Further enhancement of the field amplitude can be achieved by reducing the mode longitudinal extent, either by manipulating the light's dispersion to slow down the group velocity or, equivalently, by implementing a longitudinal resonator. Current implementations using both of these methods include quantum dots integrated into photonic crystal waveguide (PhCW) structures [6] and atoms placed in high finesse cavities [7–9].

An emerging alternative is the interaction between ultracold atoms in vacuum and slow light guided by periodic photonic nanostructures [10,11]. These hybrid systems combine the sub-wavelength confinement and dispersion control of the nanostructures with the long coherence times of isolated single atoms, all on a flexible and scalable platform. The atoms need to interact with the evanescent field around the structure. In current approaches [11–13], they are placed inside tiny 250 nm-wide slots etched into corrugated bridge waveguides. Several Bloch modes are used with frequencies tuned close to, and far from, the atomic resonance frequency for probing and trapping, respectively [2]. Although such a combined approach has a great potential, it is not without significant technical challenges. In particular, atoms need to be loaded into narrow interacting regions [14].

In this work, we outline a novel geometry for the nanostructured waveguide. By offering a $2\pi$ solid-angle access to the interaction region, it decouples the trapping and atom-field-interaction problems and allows for easier manipulation of atomic clouds or single atoms around the structure. In addition, the design takes into account two important issues which have not been considered in earlier works. First, although the evanescent field is strongest on the surface, attractive coupling of electric dipole fluctuations (Van der Waals attraction on the sub-wavelength scale) sets a lower limit on the proximity of the atoms to the nanostructure [15]. It is therefore desirable to have the interacting mode extend significantly away from the surface of the structure in order to maximize the field amplitude at a few hundreds of nanometers away from the surface. Secondly, the waveguide dispersion must be engineered so that low group velocity is achieved at a frequency that is resonant with an atomic transition. As we will see, this usually requires high accuracy in the manufactured waveguide geometry, especially for weakly modulated geometries. To ease the fabrication tolerances and lower the impact of nanostructure imperfections on slow light in general, it is therefore desirable, as shown below, to look for slow light resulting from dispersion curves with large effective photon masses (i.e. flat bands). The hybrid-clad waveguide geometry proposed hereafter represents a first attempt to fulfill both requirements.

## 2. Design requirements

**Slow light for strong coupling.** Let us first consider the interaction of atoms with translation-invariant waveguides, such as ultra-thin unclad optical fibers [1,3,16]. The strength of the interaction is characterized by the fraction $\beta$ of the atom spontaneous decay that goes into the guided mode of the waveguide: $\beta = 2\gamma_M/(2\gamma_M + \gamma')$, where $\gamma_M$ is the decay rate into the guided mode and $\gamma'$ is the decay rate into all other decay channels. A first route to achieve large coupling consists in enhancing the vacuum fluctuations associated with the guided mode to selectively increase $\gamma_M$. Alternatively, one may reduce $\gamma'$. This ambitious route that relies on the suppression of the coupling over a continuum of radiation modes has been proven successful in solid-state systems [6,17,18] only; it is mostly unworkable for atoms in vacuum since $\gamma'$ is likely to be comparable to the emission rate of the atom in free space, $\gamma_0$ [1,19]. For translation-invariant waveguides, $\gamma_M/\gamma_0 = 3/(8\pi)\, n_g\, \lambda^2/S_{\text{eff}}$ [20], where $S_{\text{eff}} = \iint \varepsilon(\mathbf{r})|\mathbf{E}(\mathbf{r})|^2 dr^2/|\mathbf{E}(\mathbf{r}_A)|^2$ [21], $\mathbf{r}_A$ denotes the atom position, $\varepsilon(\mathbf{r})$ the waveguide relative permittivity, and the atom is assumed to be in vacuum. This clearly indicates that the interaction is driven by two main quantities: the group index $n_g$ and the effective photonic mode area $S_{\text{eff}}$. Consistently with [1], we have calculated for the D$_2$ line of Cesium atoms and a 400-nm diameter fiber in vacuum, that the $\beta$-factor is only 0.07 at the fiber surface. In comparison, a photonic structure that would maintain the same effective mode area and additionally increase the group index up to $n_g \approx 50$ would reach $\beta \approx 0.79$ (a $\beta \sim 0.5$ for a group index of $n_g \sim 11$ has been recently reported [22]). One immediately sees the promising potential of periodic photonic waveguides that operates in the slow light regime [11,14,19]. Note that metallic waveguides such as chains of metallic nanoparticles that offer deep sub-wavelength transverse confinement [23] may also be of high potential interest.

**Effective mass for robustness.** Now that we have argued the necessity to implement slow light to achieve strong photon-atom couplings ($\beta \approx 1$), let us go one step further and consider design requirements imposed by the need to operate over a slow-light-frequency range that is as large as possible to relax the constraints on fabrication accuracy required to precisely match the atom transition frequency $\omega_A$. Figure 1 shows the target photonic dispersion curve $\omega(k)$ with a green-solid curve, for which a low group velocity $v_0 = d\omega/dk$ is theoretically achieved at $\omega = \omega_A$. For the sake of illustration, the dispersion curve is plotted for the hybrid-clad waveguide (*L* = 240 nm) that will be studied in detail in Section 3.

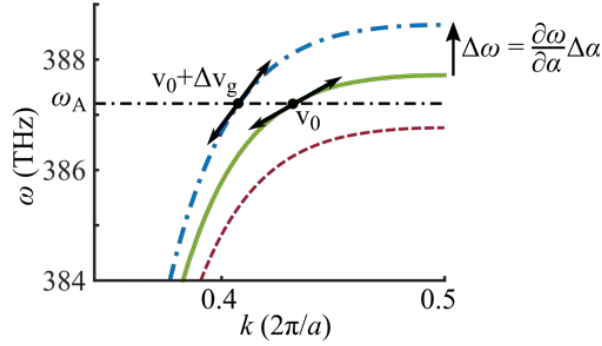

FIG. 1. Robustness to fabrication inaccuracies near the boundary of the first Brillouin zone. The green-solid curve shows the ideal (no fabrication error) dispersion curve of a hybrid-clad waveguide with a channel width of $L$ = 240 nm. At the atom-line frequency $\omega_A$, the theoretical group-velocity is $v_0$. Because of slight inevitable fabrication errors, the targeted ideal dispersion curve is never implemented, and the actual dispersion curves correspond to slightly shifted versions of the ideal dispersion curve. The blue dash-dot/red dashed curves are obtained by assuming that the channel widths are $L$ = 240 ∓ 2 nm. Thus the actual group velocity is no longer $v_0$ at $\omega = \omega_A$, but $v_0 + \Delta v_g$. Note that a blue (resp. red) shift results in a Bloch state with a larger (resp. smaller) group velocity. Note also that if the targeted group velocity is not large enough, a red shift may result in a Bloch state with an evanescent character (as illustrated in the figure).

Photonic components, such as waveguides, are never manufactured to their nominal specifications. The blue dash-dot curve in Fig. 1 represents a real version of the dispersion curve which is vertically blue-shifted away from the target dispersion curve by $\Delta\omega = (\partial\omega/\partial\alpha)\Delta\alpha$, potentially because of a tiny inevitable fabrication deviation $\Delta\alpha$ of the waveguide parameters. Consequently the actual group velocity is shifted from its target value by $\Delta v_g = (\partial v_g/\partial\omega)(\partial\omega/\partial\alpha)\Delta\alpha$ and assuming a quadratic dispersion relation close to the band edge $\omega_e$, $\omega - \omega_e = -(k - \pi/a)^2/2m$, we have

$$\Delta v_g = \sqrt{2(\omega_e + \Delta\omega - \omega_A)/m} - \sqrt{2(\omega_e - \omega_A)/m}, \tag{1}$$

where $m = (d^2\omega/dk^2)^{-1}$ is the effective photon mass of the dispersion curve, the second derivative being calculated at $\omega = \omega_e$. Equation (1), shown with the red curve in the inset, also represents small group-velocity mismatches experienced by light as it propagates in a real waveguide with an inevitable residual disorder that is due to fabrication imperfections. For a vertically red-shifted dispersion curve, as illustrated with the red dashed curve in Fig. 1, an equation similar to Eq. (1) with an inverse square-root dependence on the mass can be derived. It is important to note that, if the targeted group velocity $v_0$ is not sufficiently large, a red shift may result in a Bloch state with an evanescent character (as illustrated in Fig. 1) at the atom transition frequency $\omega_A$, which would be very detrimental [24]. Hence, red-shifts impose a minimum value for the targeted group velocity

$$v_0 > \sqrt{2\Delta\omega/m}, \tag{2}$$

which also scales as the inverse square-root of the mass. Although illustrated here with a simple frequency shift of the band, the inverse-square-root argument is general. It underlines that operation at predetermined small group velocities imposes strong requirements on the quality of the fabrication processes and that the effects of fabrication shifts or fluctuations can be lowered by selecting photon modes with large effective mass (i.e. flat bands), typically obtained with large index modulations. Let us note that the frequency shift $\Delta\omega = (\partial\omega/\partial\alpha)\Delta\alpha$ is difficult to lower by design since close to the band edge, these shifts are directly proportional to variations of the waveguide effective index $n_{eff}$, $\Delta\omega/\omega = \Delta n_{eff}/n_{eff}$.

**Photonic crystal versus Bragg waveguides.** In Bragg-type periodic waveguides, guidance is achieved by total internal reflection on the waveguide lateral clad walls and light slowdown results from multiple reflections due to the periodicity. Examples of Bragg-type waveguides are sub-wavelength periodic chains of holes or boxes [25], or corrugated-slot (or "alligator") waveguides [11,14,22]. Such index-guided waveguides, see Fig. 2a, operate much like one-dimensional Bragg stacks. The effective mass is proportional to the refractive index modulation [26,27] which is in general quite small with dielectric structures. In addition, for guided modes that extend out of the waveguide into the air clad as required for interaction with distant atoms, the mode field only weakly sees the modulation and thus the effective index modulation and the effective mass become very small. Clearly, TIR guidance leads to sharp frequency matching conditions that are difficult to guarantee with the fabrication imperfections.

In contrast, for photonic crystal waveguides, guidance is due to light reflection on a photonic gap, see Fig. 2b, and by tuning some geometrical parameters, the phase $\phi$ of the reflection coefficient can be made highly dispersive [28]. As a result (Fig. 2c), the dispersion curve and the group-velocity-dispersion can be engineered even for small group velocities [29], and the effective mass may take values that are out of the reach of Bragg-type periodic waveguides [26].

Figures 2d-2e compare the dispersion relation of a few relevant waveguides. In (d), we consider the alligator geometry already used in cold-atom experiments [11,14,22] and the hybrid-clad waveguide that is studied in Section 3. For the sake of completeness, we also provide in (e) more classical periodic waveguides, e.g. the single-row missing photonic-crystal waveguide (with an hexagonal lattice constant a = 420 nm, hole radius 0.3a and an effective index of 2.83 to model the transverse confinement of the main TE-mode in a silicon membrane of thickness 240 nm suspended in air [30]) and the 1D air-semiconductor Bragg stack (composed of alternating dielectric layers with lattice constant a = 453 nm and refractive indices $n_1$ = 1.5 and $n_2$ = 3.5). The computational results illustrate our purpose well, since we find that the masses of the alligator waveguide ($mc$ = 0.6 µm$^{-1}$) and 1D Bragg stack ($mc$ = 4 µm$^{-1}$) are substantially smaller than those of the hybrid-clad waveguide ($mc$ = 36 µm$^{-1}$) and photonic-crystal waveguide ($mc$ = 40 µm$^{-1}$).

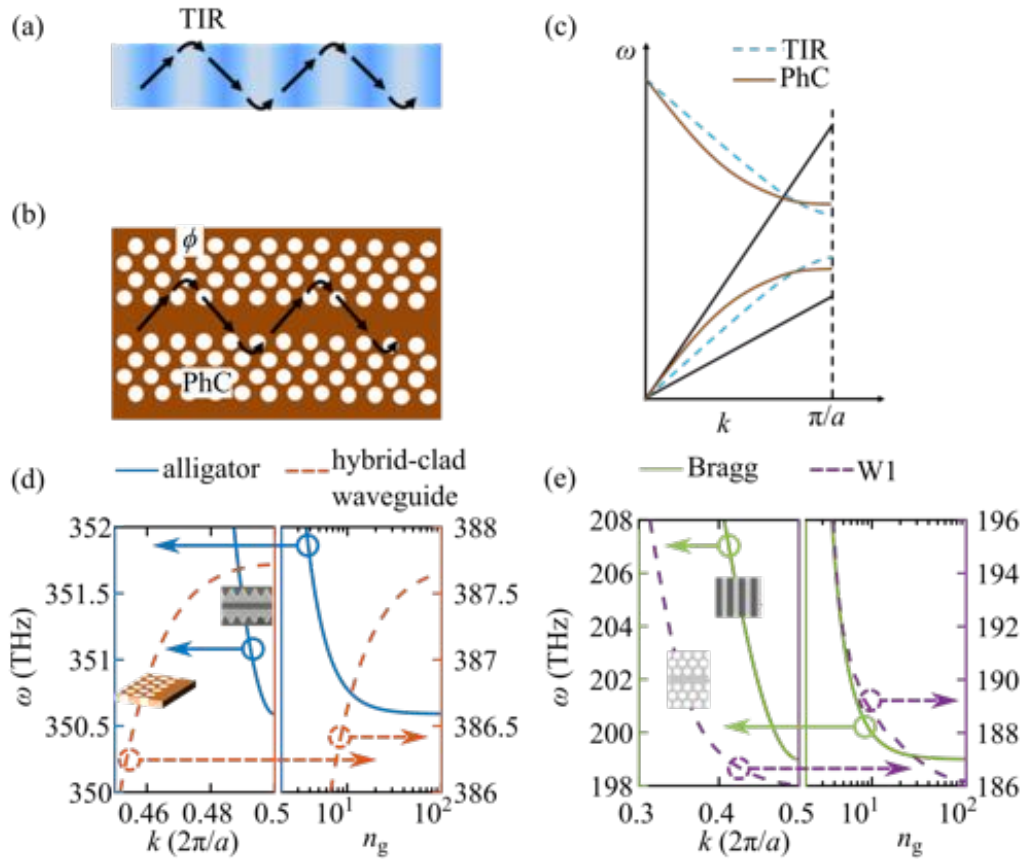

FIG. 2. TIR versus PhC guidance for controlling mode flatness. (a) Guidance is achieved with total-internal-reflection (TIR) in Bragg-type waveguides. The effective index modulation is very weak, especially for extended modes. This results in very small effective mass. (b) With photonic crystal (PhC) guidance, the zig-zag propagation exhibits an additional phase $\phi$ at the reflection on the PhC clads, and $d\phi/d\lambda$ represents an additional degree of freedom to control the mode flatness. (c)

Schematic dispersion curves of structures shown in (a) and (b). (d) Dispersion curves of alligator (solid-blue, $mc = 0.6$ µm$^{-1}$) waveguide [11] and hybrid-clad waveguide for $L = 240$ nm (dashed-red, $mc = 36$ µm$^{-1}$). (e) Dispersion curves of 1D-Bragg waveguide (green solid, $mc = 4$ µm$^{-1}$) and W1 waveguide (purple dashed, $mc = 40$ µm$^{-1}$) [26].

An alternate reasoning would have consisted in arguing that the impact of fabrication errors is lowered with dispersion curves that offer vanishing second-order dispersions at the operating frequency $\omega_A$, implying that $v_g \approx v_0$ on a small spectral interval around the operating frequency $\omega_A$. Essentially, this approach leads to the same conclusion, the necessity to use PhC cladding to engineer the dispersion curve with the additional degree of freedom given by the phase dispersion $d\phi/d\lambda$ at the reflection.

## 3. Hybrid-clad waveguides

The proposed hybrid-clad waveguide, see Fig. 3(a), combines the two guidance mechanisms. On one lateral side of the waveguide, the fundamental mode is gap-guided by a periodic array of holes, which allows us to control the flatness of the dispersion band, and on the opposite lateral side, the mode is index-guided by a sharp vertical sidewall, which offers an entire half space to host the electric field extended into the vacuum cladding. The geometry leaves many options available to manipulate and trap atoms in this area including free space optical tweezers, magnetic trapping using static magnetic fields possibly integrated into a chip structure with the waveguide, and bi-chromatic dipole trapping with other low-energy mode of the hybrid waveguide.

The design faces two conflicting perspectives. On one side, to implement an atom-photon interaction that occurs away from the nanostructure, one needs to synthesize a Bloch mode with a field that is well extended into the vacuum cladding. This implies that the Bloch mode weakly interacts with the periodic structure. On the other side, to control slow-light flatness, one needs a strong interaction between the Bloch mode and the periodic modulation, keeping away from any weak-permittivity-modulation regime for which the group velocity of the Bloch mode and the band flatness are hardly controllable. Thus the design appears as an exercise of compromise. In the reminder of the paper we present the design of a promising example of this trade-off.

We assume that the waveguide is etched into a silicon nitride (SiN) membrane. Photonic crystal structures based on SiN and fabricated with electron-beam writers and dry etching have already been manufactured for visible-light applications with great accuracy [31,32]. The photonic crystal is formed by cylindrical holes perforated in the membrane on a triangular lattice. The propagation direction $z$ coincides with the Γ-K direction of the crystal. The waveguide thus depends on four parameters, the membrane thickness $h$, the lattice period $a$, the hole radius $r$ and the channel width $L$, as shown in Fig. 3(a). We assume a typical thickness of 300 nm.

The dispersion curve is calculated with a mode-solver initially developed for studying light propagation in photonic crystal waveguides operating above the light cone [33] and further refined to study light scattering in periodic waveguides [34]. The solver relies on an analytical integration of Maxwell's equations along the direction of periodicity ($z$ axis), and on a supercell approach with real, nonlinear coordinate transforms (using stretched-coordinate transforms) along the two transverse directions ($x$ and $y$ axes). The transforms allow us to carefully handle far-field radiation conditions, even when the guided mode extends away in the vacuum cladding. The supercell includes 9 rows of holes on one side of the waveguide and a vacuum gap with a 1-µm-width on the other side, for a total area of 5.4 × 1.4 µm$^2$. The numerical sampling in the transverse directions is performed in the Fourier domain using truncated Fourier series [33,34]. All numerical results are obtained by retaining a total number of 51×29 Fourier harmonics, but other calculations performed for larger truncation orders have revealed that the truncation error has no influence on the following analysis.

The first step of the design consists in finding the range of photonic crystal parameters $a$ and $r$, for which the photonic crystal membrane offers a TE–like bandgap for the Rubidium D$_2$ transition line at $\lambda_{Rb} \approx 780$ nm. This initial calculation was performed with the MIT Photonic-Bands (MPB) package [35]. In the second step, we choose the crystal parameters in the bandgap range and we tune the light-channel width $L$ to shift the band edge of the dispersion curve close to the light line, where the modes tend to be significantly delocalized from the waveguide. This step is pivotal to obtain a slow Bloch mode with a field distribution that penetrates far into the vacuum cladding. As $L$ is reduced, the dispersion curve shifts towards high frequencies; however as the band-edge approaches the vacuum light line, we observe that the effective mass drops at a fast rate, see the mass values in the caption of Fig. 3, as if the mode couples to the continuum of radiation states at slow group velocities. We have performed several tries, repeating the second step for different crystal parameters of $a$ and $r$ in the bandgap range computed with the MPB package.

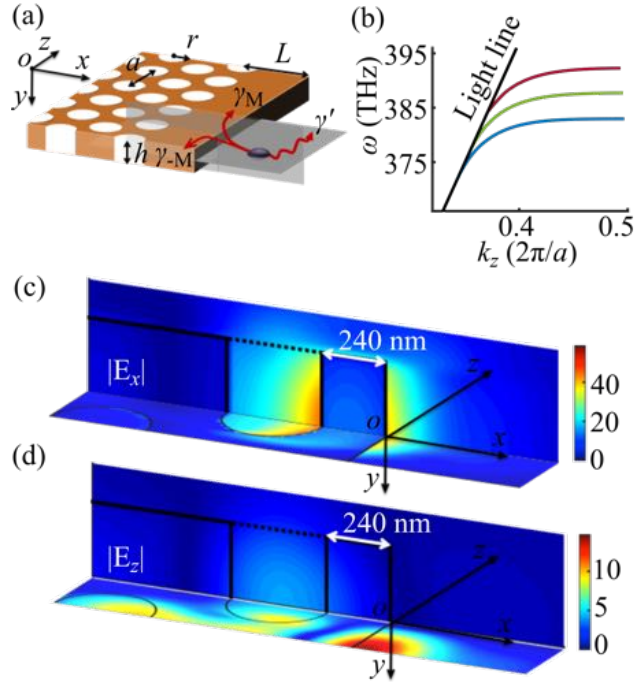

FIG. 3. Hybrid-clad waveguide geometry. (a) Sketch of the geometry (only four rows of air holes are shown on the left side of the waveguide). The atoms decay into the fundamental guided mode and the continuum of radiation modes with decay rates $\gamma_M$ and $\gamma_{\text{rad}}$, respectively. (b) The red, green and blue (from top to bottom) dispersion curves represent the guided TE-like Bloch mode for channel widths, $L$ = 230 nm, 240 nm and 250 nm and effective masses $mc$ = 22.6, 36.1 and 177.6 µm$^{-1}$, respectively. The band gap of the bulk photonic crystal opens in a frequency range of [345, 400] THz. The vacuum light line is shown with the black line. (c-d) Distributions of the dominant electric-field components ($|E_x|$, $|E_z|$) for $n_g$ = 50 along the grey symmetry planes shown in (a). All results in (b)-(c) hold for $a$ = 340 nm, $r$ = 0.3$a$, and $h$ = 300 nm.

The curves in Fig. 3(b) shows the dispersion curve of the TE-like mode of the hybrid-clad waveguide for $a$ = 340 nm, $r$ = 0.3$a$, $h$ = 300 nm, and $L$ = 230, 240 and 250 nm. For this set of parameters, the hybrid-clad waveguide is monomode in the band gap. The corresponding photon effective masses $mc$ are 22.6, 36.1 and 177.6 µm$^{-1}$, respectively. For comparison, we note that the mass of the alligator waveguide, fitted from numerical data extracted from Fig. 1b in [11], is only 0.6. Figure 3(c) displays the distributions of the dominant electric-field components in the symmetry planes of the hybrid-clad waveguide for $L$ = 240 nm (note that the $y$-component of the electric field is null in the $x$-$z$ plane for TE-like modes). The calculation is performed at a frequency corresponding to a group index of 50, but very similar mode profiles are obtained for other group indices, $n_g$ = 10-100. What is remarkable in the plot is that the electric-field distribution is primarily intense in vacuum, in the first and second rows of holes and more importantly it is even more intense in the vacuum cladding on the right side of the hybrid-clad channel. Indeed, a large fraction of the mode energy is in vacuum, as evidenced by the value of the effective index (= 1.09) that is very close to one. This observation holds for all electric-field components. Additionally, if we pay attention to the light-blue areas rather than to the red intense spots, we note that the mode extends quite far away above the membrane and on the right side of hybrid-clad channel, indicating that strong atom-photon coupling at 200-300-nm transverse separation-distances is possible.

## 4. Atom-field coupling

An important figure of merit [11,19] that characterizes the strength of the atom-photon coupling is the single-atom reflectivity, which is, at zero detuning and low saturation condition, directly linked to the $\beta$-factor. The single-atom reflectivity can be defined as the reflection of a coherent multiphoton laser field at frequency $\omega_L$ launched into the fundamental Bloch mode of the waveguide and illuminating a single atom initially in its ground state. During the interaction, the incident photons first polarize the atom with an induced dipole moment that is proportional to the driving Bloch-mode field-vector $\tilde{\mathbf{E}}(\mathbf{r}_A, \omega_L)$ at the atom position $\mathbf{r}_A$, and then the atom radiates by exciting the guided modes and all other radiation modes with decay rates $2\gamma_M$ and $\gamma'$.

This scattering problem has been recently analyzed with a fully-quantum treatment based on a combination of electromagnetic Bloch-mode-expansion and Green-tensor techniques with the optical Bloch equations for the atom density-matrix operator. In the weak coupling regime and for a two-level atom, it was shown that the Bloch-mode reflection coefficient is given by [19]

$$r = \sigma_0 \frac{-j\varepsilon_0 c}{4\mathcal{P}} \frac{(2\delta/\gamma_0 - j\gamma/\gamma_0)}{4(\delta/\gamma_0)^2 + 2|\Omega/\gamma_0|^2 + (\gamma/\gamma_0)^2} \tilde{\mathbf{E}}(\mathbf{r}_A, \omega_L) \cdot \tilde{\mathbf{E}}(\mathbf{r}_A, \omega_L), \tag{3}$$

where $\delta = \omega_L - \omega_A$ is the detuning between the laser driving frequency and the atom resonance frequency, $\mathcal{P}$ the power flow of the normalized incident Bloch-mode ( $\mathcal{P}$ = 1), $\sigma_0 = 6\pi c^2/\omega_A^2$ the extinction cross section on resonance of an isolated quantum mechanical two-level system, $\varepsilon_0$ the permittivity of vacuum, $c$ the light speed, $\Omega$ the complex external Rabi frequency which takes into account possible saturation effects and $\gamma = 2\gamma_M + \gamma'$ is the total decay rate in all the photon modes. The challenging part for calculating $r$ is the calculation of the total decay which necessitates the evaluation of the coupling into radiation modes. The exact calculation can be performed with Bloch-mode expansions [6,17]. However by assuming that $\gamma' = \gamma_0$, the total decay rate takes a simple expression $\gamma/\gamma_0 \approx 1 + (3\pi c/2\omega^2)|\tilde{\mathbf{E}}(\mathbf{r}_A, \omega_L)|^2$, and the reflection coefficient, which is slightly underestimated, is straightforwardly calculated for any driving laser frequency and for any atom location. As shown in [19], this approximation is valid and leads to accurate results for the small group velocities.

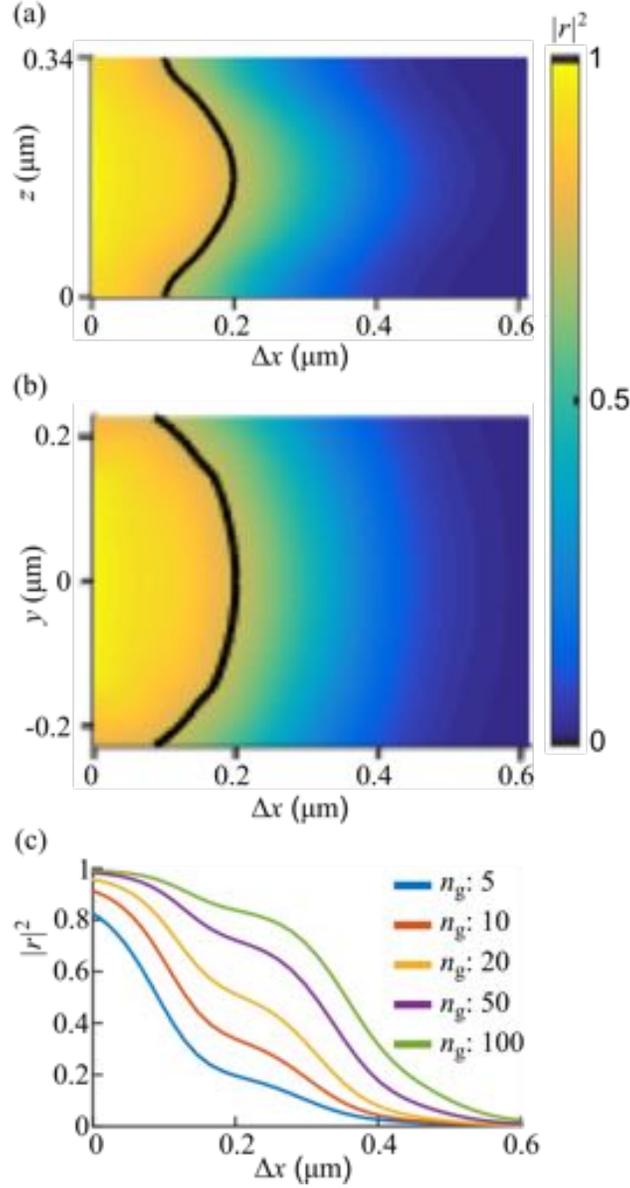

FIG. 4. Reflectance of a single atom for $\omega_L = \omega_A$ as a function of the atom position in the main symmetry planes. $\Delta x$ denotes the $x$-separation distance from the vertical sidewall of the waveguide ($\Delta x = 0$ corresponds to the sidewall boundary). (a) The atom is placed in the $x$-$z$ horizontal plane, see Fig. 3a, and $n_g = 50$. (b) The atom is placed in the $x$-$y$ vertical plane, see Fig. 3a, and $n_g = 50$. (c) Reflectance for $y = 0$ and $z = a/2$ and for $n_g = 2, 10, 50, 100$. The black contour line in (a) and (b) indicates a reflectance of 0.75.

In Fig. 4, we display the reflectance distribution at zero detuning and low saturation ($|\Omega/\gamma_0|^2 \ll (\gamma/\gamma_0)^2$) as a function of the atom position in the main symmetry planes. Figure 4a is obtained for the $x$-$z$ horizontal median plane (see Fig. 3a), in which only the $x$- and $z$-components of the electric-field are involved in the atom-waveguide coupling, and Fig. 4b for the $x$-$y$ vertical plane (see Fig. 3a), in which all field components are involved. Owing to the diminishing group velocity, $n_g = 50$, large reflectance values are obtained when the atomic line frequency is matched with the photonic resonance. Importantly, we note that strong atom-photon couplings are predicted for large separation distances $\Delta x$ from the vertical sidewall (the black contour line indicates a reflectance $|r|^2 = 0.75$). Figure 4c shows the reflectance as a function of the $x$-coordinate for $y = 0$ and $z = a/2$. The results are shown for several group velocities, $n_g = 2, 10, 50$ and $100$. The resonant reflectance for $n_g = 50\text{-}100$, is about 0.6-0.8 for atoms trapped at a separation distance of $\Delta x = 300$ nm from the waveguide, implying that $\beta$ is as large as 0.8-0.9. Such a

strong atom-photon coupling far from any waveguide surface is unprecedented in atomic/molecular optical physics and could create new opportunities for high precision studies.

## 5. Discussion and conclusions

By design, the present hybrid-clad waveguides are responding to many important challenges, such as atom loading, light scattering from device imperfections, atom-photon resonance matching and robustness to vacuum-force fluctuations. Their fabrication for operation at $\lambda_{Rb} \approx 780$ nm appears achievable, in view of recent achievements of the photonic-crystal community, either with silicon nitride [32] or gallium phosphide [36] membranes.

Another important issue to address is how to inject light into the waveguides from focused beams or lensed fibers. We believe that efficient injection can be achieved in three stages. First light from a fiber can be coupled into a single-mode nanowire with very small insertion loss (< 1 dB) [37,38]. Then the nanowire mode can be tapered into the fast guided mode of hybrid-clad waveguide. At this stage to avoid any group-velocity mismatch problems, one may couple into a hybrid-clad waveguide with a reduced width $L$. Efficient adiabatic mode transformations between fast modes can be achieved with tapers implementing an effective index gradient through a progressive variation of hole dimensions [39]. Due to the large group velocity, the field does not expand much in the photonic-crystal mirror so that the mirror dominantly acts as a step barrier (TIR-like guidance), and the asymmetry of the hybrid-clad waveguide is likely to be effectively managed with the progressive variation. The last step corresponds to a slow decrease of the group velocity by adiabatically increasing the hybrid-clad channel width towards the desired one. Note that very short (1-2 μm long) and very efficient tapers (90% efficiency) have been manufactured for large group-velocity impedance mismatches with photonic-crystal waveguides [30,40,41].

In general, slow light is a consequence of the interacting of modal electromagnetic fields with a periodic change of the refractive index, and efficient light slowdown is achieved when the mode fields propagate inside a corrugated structure. However, this fact conflicts with the desire for implementing flat dispersion bands with mode fields dominantly residing in vacuum claddings where the atoms are loaded. The present design that uses slow-light in photonic crystal waveguides is certainly not optimal, but it offers the benefit to directly address the conflict and to present a compromised solution. We are working toward extensions of this work with more complicated geometries by numerical simulation and device fabrication. Our efforts are motivated by the predictions in Fig. 4, which may lead to a new generation of devices offering a strong photon interaction with atoms that are trapped with negligible vacuum-force interaction.

**Acknowledgments.** XZ and PL acknowledge financial support from Université de Bordeaux and Région d'Aquitaine. RF has received financial support from the French "Direction Générale de l'Armement" (DGA). This research was supported by the UK Engineering and Physical Sciences Research Council (grant EP/E036473/1).